\documentclass[letterpaper]{jpconf}
\usepackage{graphicx}
\usepackage{wrapfig}
\usepackage{epsfig,subfigure}
\usepackage{amsmath,amstext,amssymb}
\usepackage{array}
\usepackage{longtable}
\usepackage{ulem}
\usepackage{color}
\def\etal{\textit{et al.}}
\def\Journal#1#2#3#4{{#1} {\bf #2}, #3 (#4)}

\def\PRL{Phys. Rev. Lett.}

\begin{document}
\title{Radiative/EW penguin decays at Belle}
\author{Nanae Taniguchi (for the Belle Collaboration)}
\address{KEK-IPNS, Institute of Nuclear and Particle Studies, High Energy Accelerator Research 
Organization}
\begin{abstract}
We present recent results for radiative and electroweak penguin decays of $B$ meson at Belle.  Measurements of differential branching fraction, isospin asymmetry, $K^*$ polarization, and forward-backward asymmetry as functions of $q^2$ for $B \to K^{(*)}ll$ decays are reported.  For the results of the radiative process, we report measurements of branching fractions for inclusive $B\to X_s \gamma$ and the exclusive $B\to  K \eta' \gamma$ modes.
\end{abstract}
% *****************************************
\section{Introduction}
$b \to s$ transition is the Flavor changing neutral currents (FCNC) which are forbidden at the tree level
in the Standard Model. However, loop-induced FCNC (called penguin decays) are
possible.
New particles in the loops can give effects at the same order as Standard Model contributions.
The process is a sensitive probe to new physics.
% *****************************************
\section{Analysis techniques}
$B$-factory provide large clear sample of  $\Upsilon(4S)$ decays $B\bar{B}$ pairs.
The main background source  comes from continuum events
($e^+e^-\to q\bar{q}(\gamma)$, $q=u,d,s,c$). To suppress the continuum
background, we use a selection criteria making use of the difference of
the event topology between $B$ decays and continuum events.
In the inclusive analysis, 
these continuum backgrounds are subtracted using the off-resonance data
sample taken slightly below the $\Upsilon(4S)$ resonance.
In the exclusive measurements, one can require the kinematic constraints
on the beam-energy constrained mass $M_{\rm bc} = \sqrt{E^*_{\rm beam} -
p^*_B}$  and
$\Delta E = E^*_B - E^*_{\rm beam}$, using the beam energy $E^*_{\rm
beam}$ and momentum $p^*_{B}$ and $E^*_B$ of $B$ candidate in the
center-of-mass system (c.m.s).
\section{$B \to K^{(*)}ll$}
The decay $b \to sll$ is induced through penguin or box diagrams at lower order\cite{bib:bsll}.
There are many observable such as branching fraction, isospin asymmetry and  forward-backward asymmetry where new physics can contribute.
These observable can be interpreted in term of Wilson coefficients.
Three Wilson coefficients, $C_{7,9,10}$ contribute.
The {\cal B}($B \to X_s \gamma$) can constraint to $|C_7|$.
The $b \to s ll$ is sensitive to sign of $C_7$.

We have measured $b \to sll$ exclusively ($B \to K^{(*)}ll$) on 657M $B\bar{B}$ pairs~\cite{bib:bsll_belle}.
10 final state ($K^+\pi^-, K_s \pi^+, K^+\pi^0, K^+$ and $Ks$) are reconstructed for $K^{(*)}$ and combined with electron and muon pairs.
$B$ meson is exclusively reconstructed with $M_{\rm bc}$ and $\Delta E$.
Main backgrounds are continuum event  and semi-leptonic $B$ decays.
The continuum background is suppressed using information of event topology and the semi-leptonic $B$ decays are suppressed using information of missing mass and  lepton vertex separation.
Dominant peaking background from $B \to J/\psi(\to ll)X$ and $\psi(2S)(\to ll)X$ decays are rejected in the $q^2$(invariant mass of dilepton).
 
\begin{figure}[h]
\vspace{-20pt}
\begin{center}
\includegraphics[width=9cm]{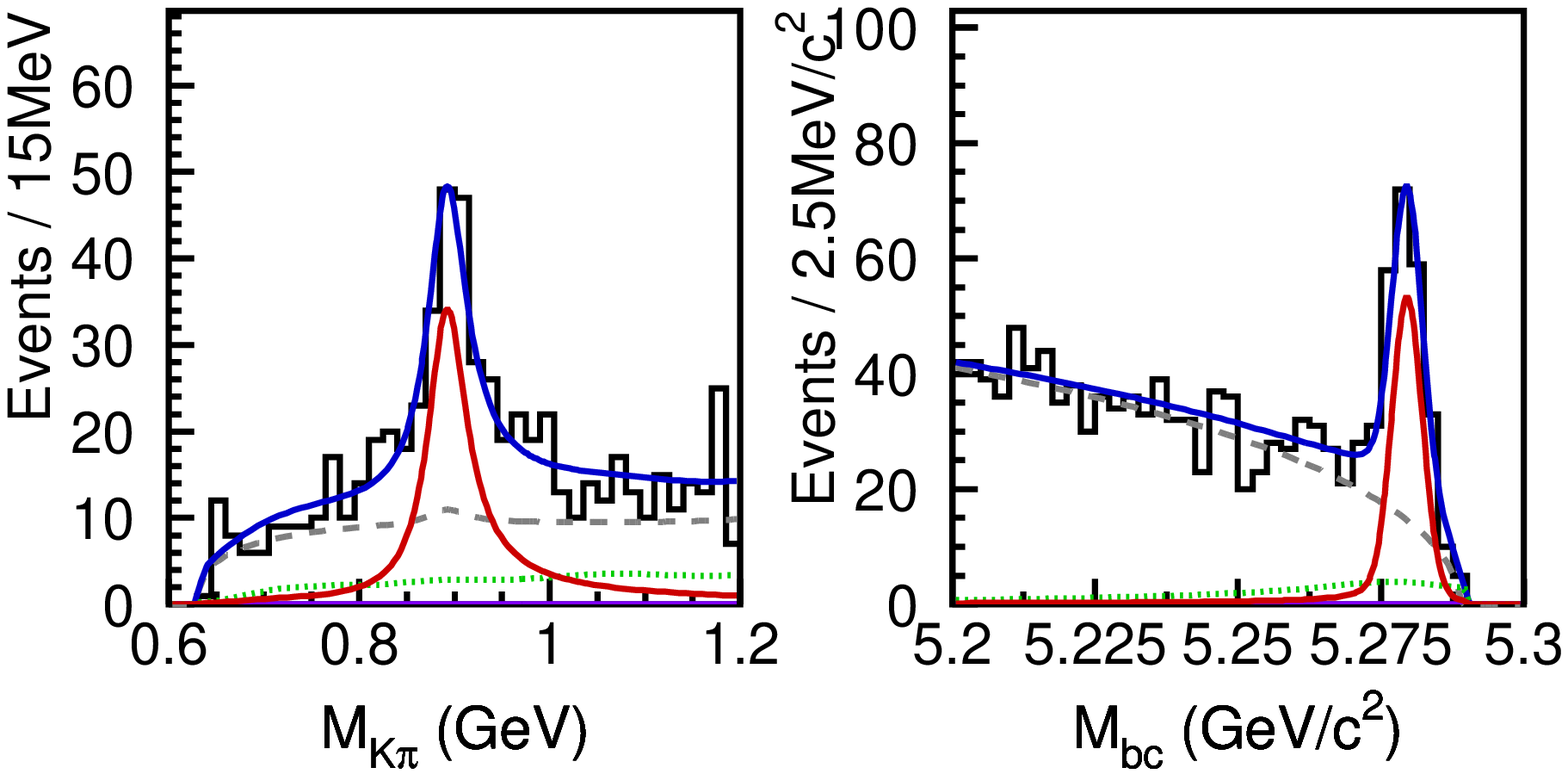} 
\hspace{10pt}
\includegraphics[width=4.5cm]{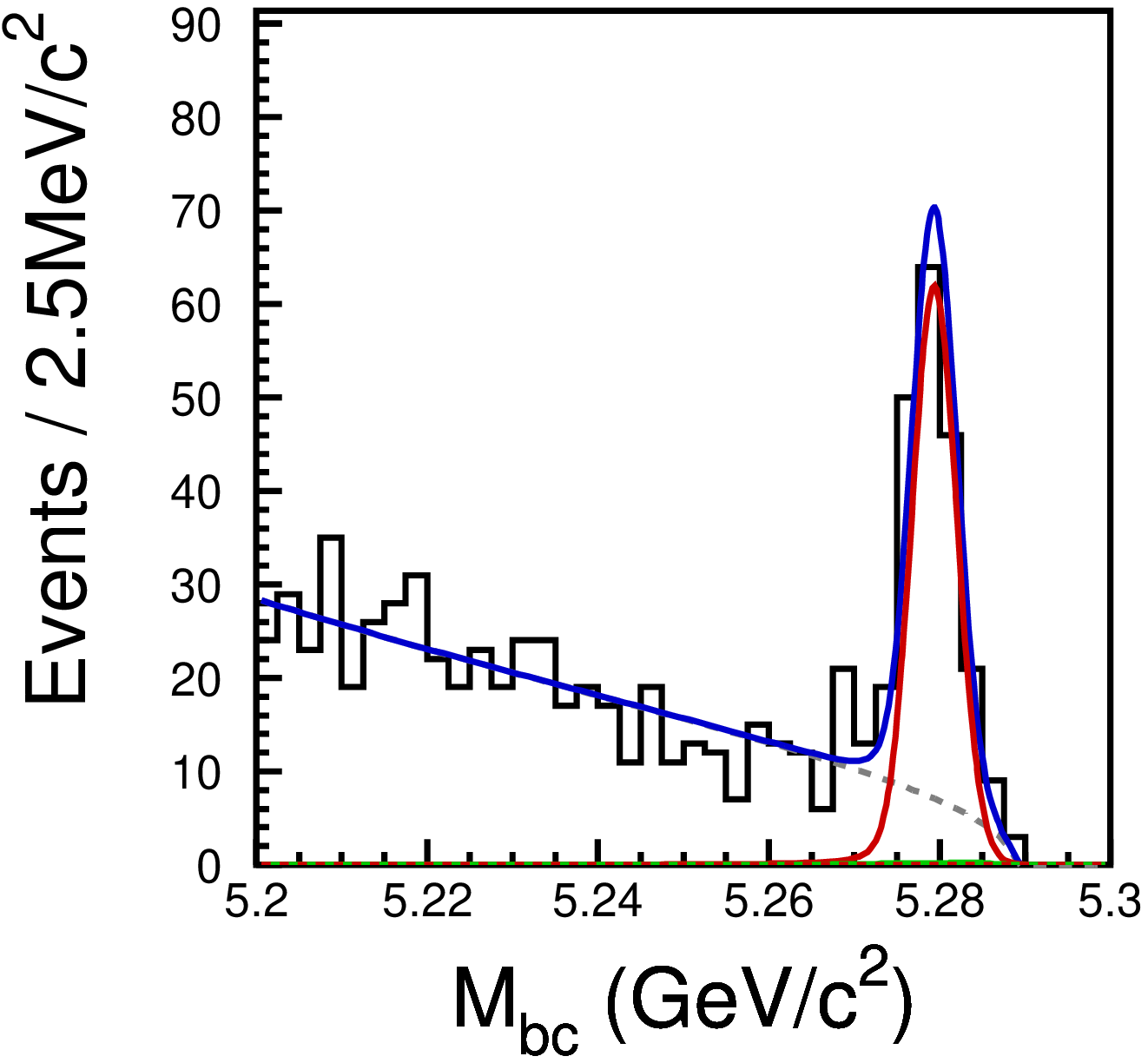}
\caption{
Distributions of $M_{K\pi}$ ($M_{\rm bc}$) with fit results superimposed for
the events in the $M_{\rm bc}$ ($M_{K\pi}$) signal region.
The solid curves, solid peak, dashed curves, and dotted curves represent 
the combined fit result, fitted signal, combinatorial background, and $J/\psi(\psi^\prime) X$ background, 
respectively.
}
\label{fig:bf}
\end{center}
\end{figure}

We obtain ${\cal B}(B \to K^* ll)=(10.8 \pm 1.0 \pm 0.9)\times 10^{-7}$ and  ${\cal B}(B \to K ll)=(4.8{}^{+0.5}_{-0.4}  \pm 0.9)\times 10^{-7}$ by fitting to $M_{\rm bc}$ (and $M_{K\pi}$ for $K^*ll$).
Fig.~\ref{fig:bf} shows the distributions of $M_{K\pi}$ ($M_{\rm bc}$) with fit results superimposed for the event in
 the $M_{\rm bc}$ ($M_{K\pi}$) signal region.
\vspace{5pt}

\begin{figure}[h]
\vspace{-5pt}
 \begin{minipage}{0.5\hsize}
\begin{center}
\includegraphics[width=4cm]{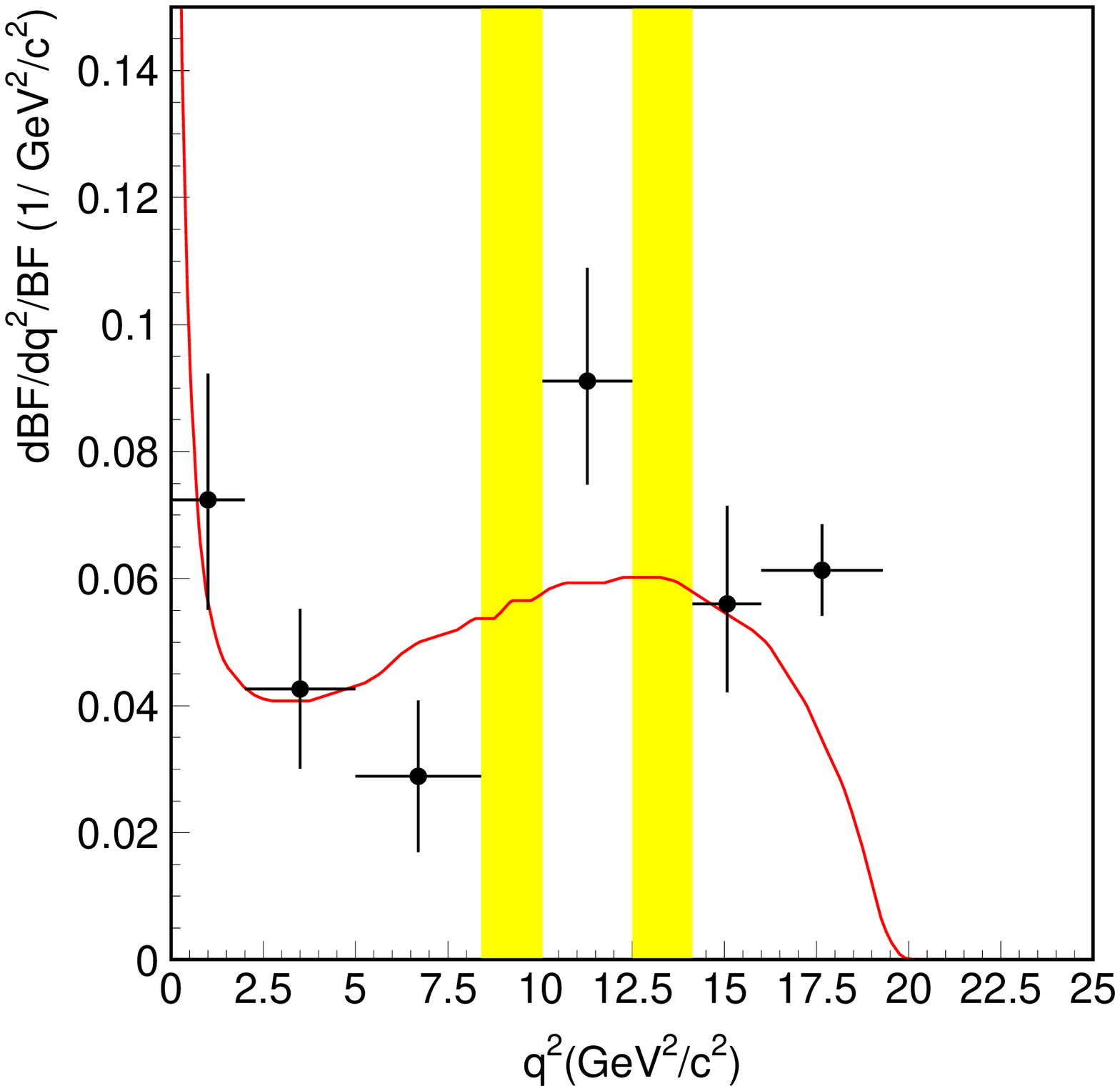}
\hspace{-10pt}
\includegraphics[width=4cm]{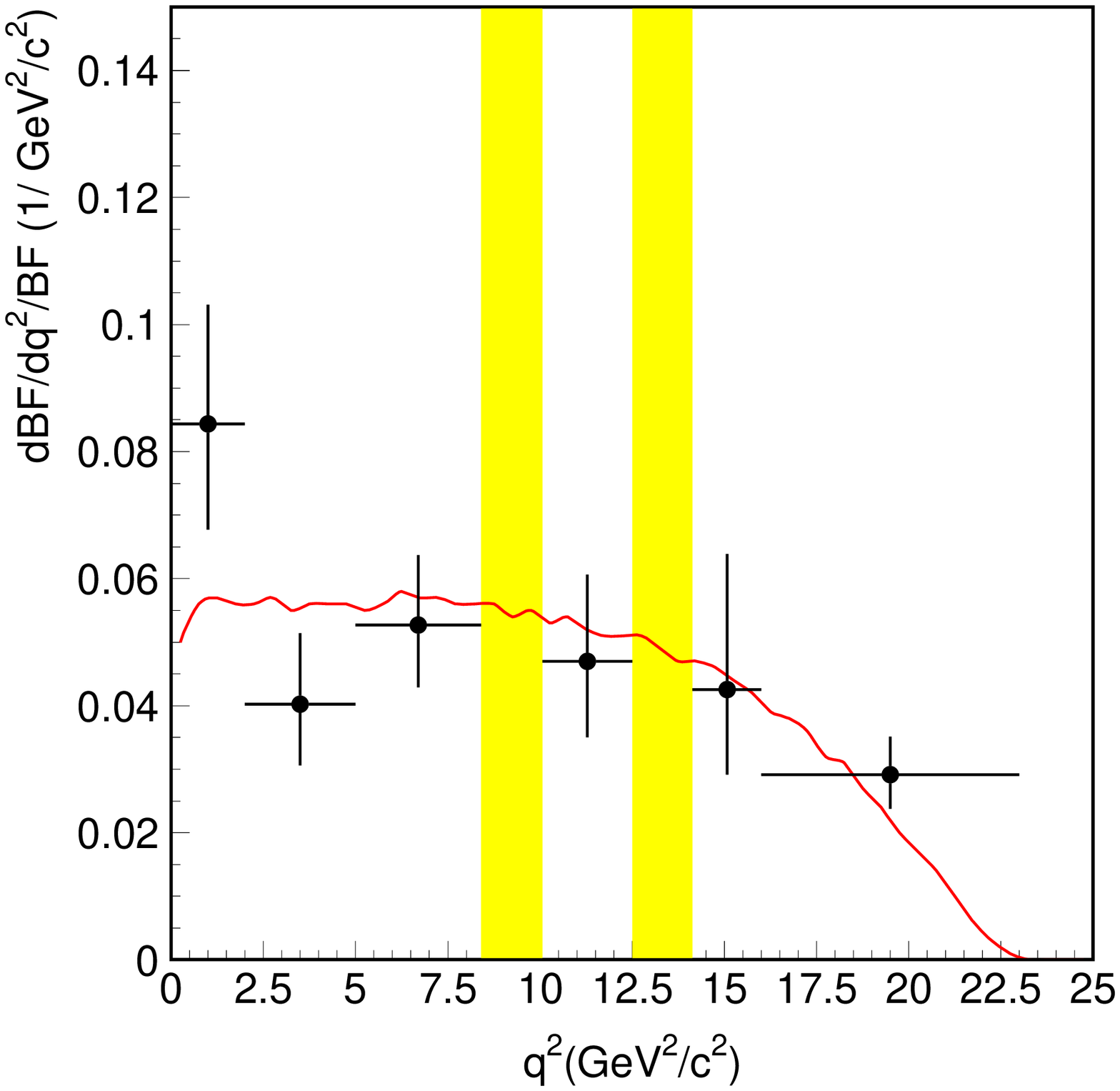}
\caption{
Differential branching fractions for 
 $K^* \ell^+ \ell^-$ (left) and $K \ell^+ \ell^-$ (right) modes as a function of $q^2$. 
The two shaded regions are veto windows to reject $J/\psi(\psi^\prime) X$ events. 
The solid curve is the theoretical prediction~\cite{Ali:2002jg}.
}
\label{fig:dbf}
\end{center}
\end{minipage}
\hspace{20pt}
 \begin{minipage}{0.5\hsize}
\begin{center}
\includegraphics[width=7cm]{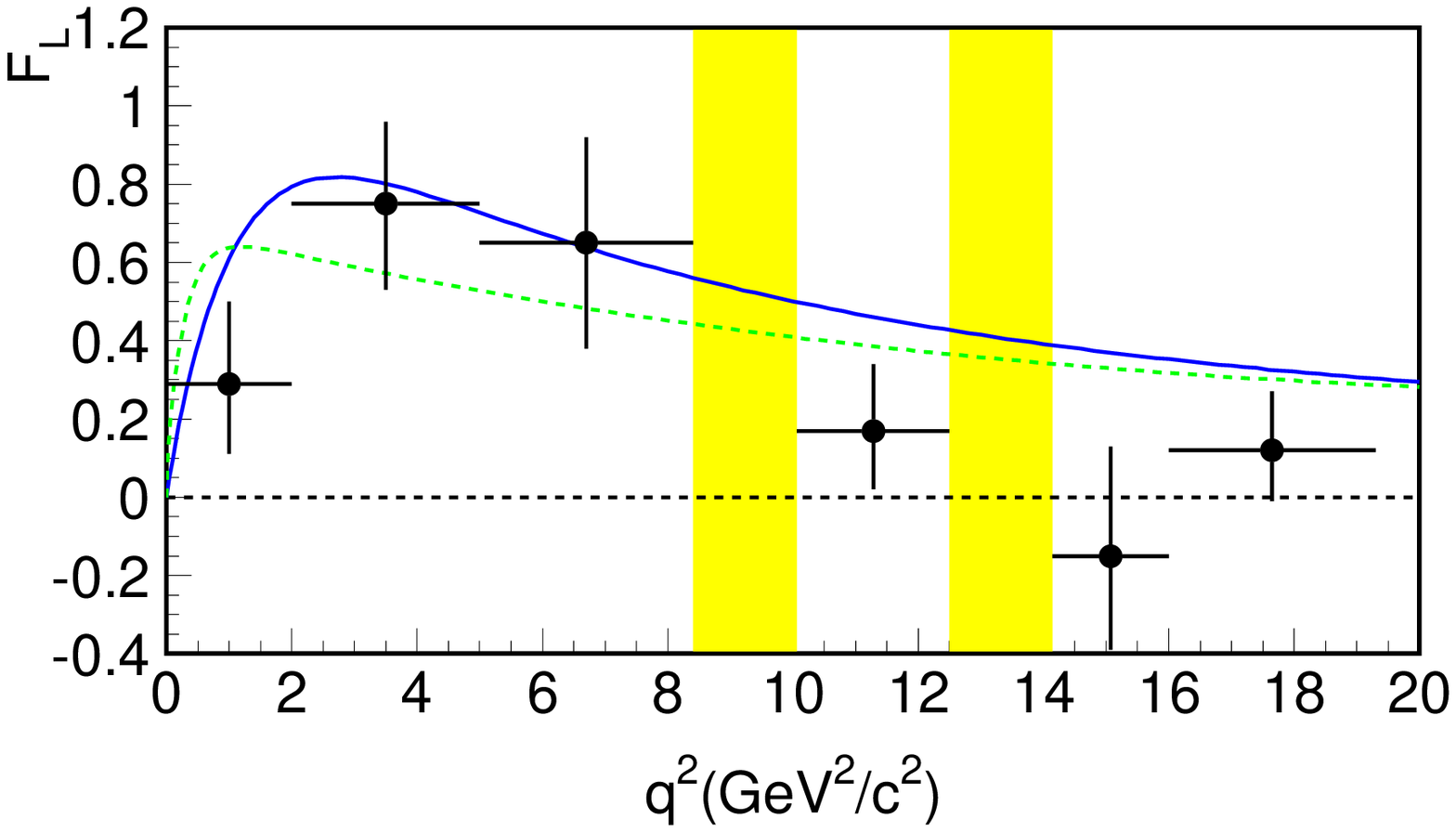}
\caption{
Fit results for $F_L$ as a function of $q^2$. 
The solid (dashed) curve shows the SM ($C_7=-C^{SM}_7$) prediction.
}
\label{fig:fl}
\end{center}
\end{minipage}
\end{figure}

\begin{figure}[h]
\begin{minipage}{0.5\hsize}
\begin{center}
\includegraphics[width=7cm]{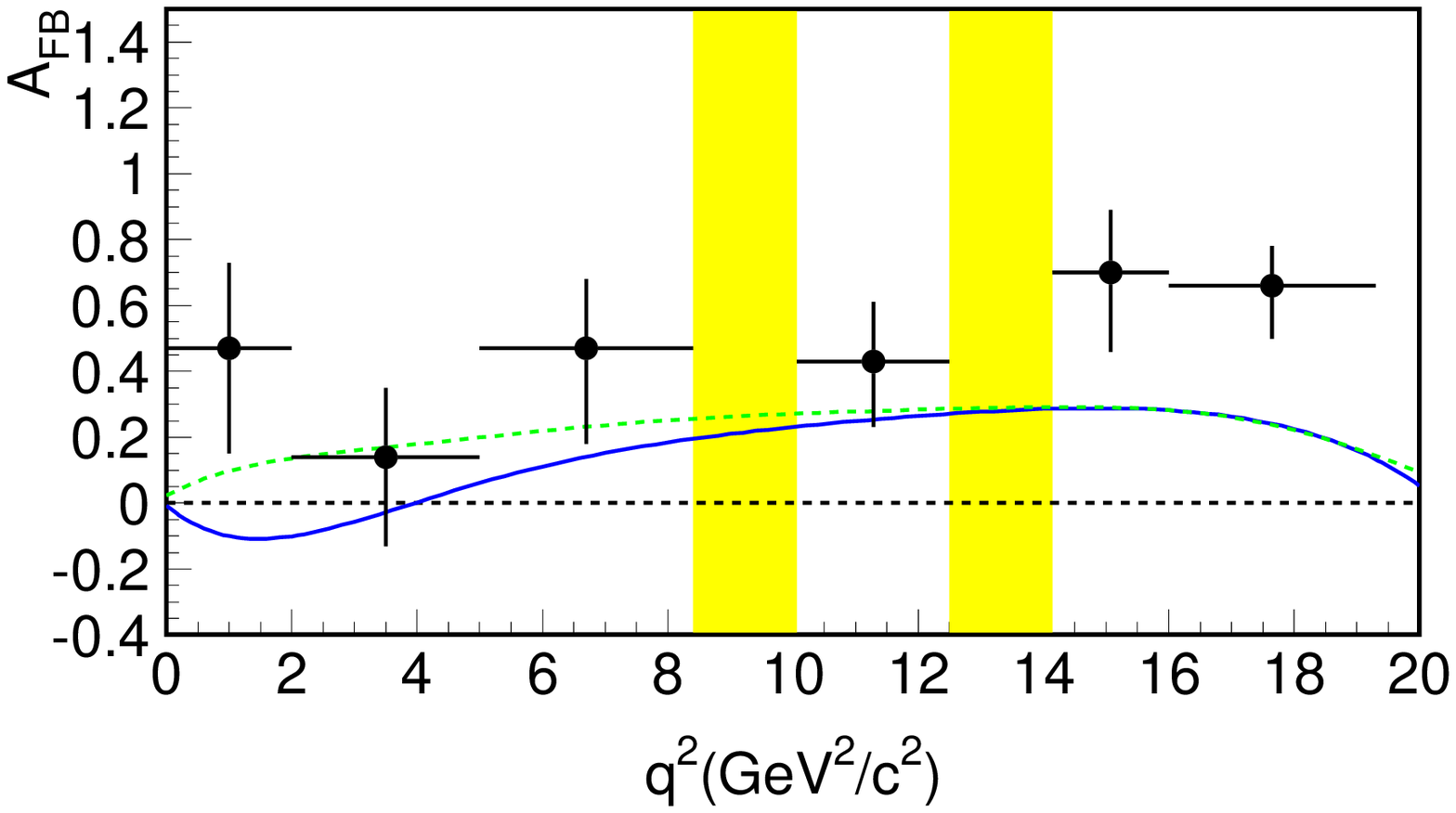}
\caption{
Fit results for $A_{FB}$ as a function of $q^2$.
The solid (dashed) curve shows the SM ($C_7=-C^{SM}_7$) prediction.
}
\label{fig:afb}
\end{center}
\end{minipage}
\hspace{10pt}
\begin{minipage}{0.5\hsize}
\begin{center}
\includegraphics[width=9cm]{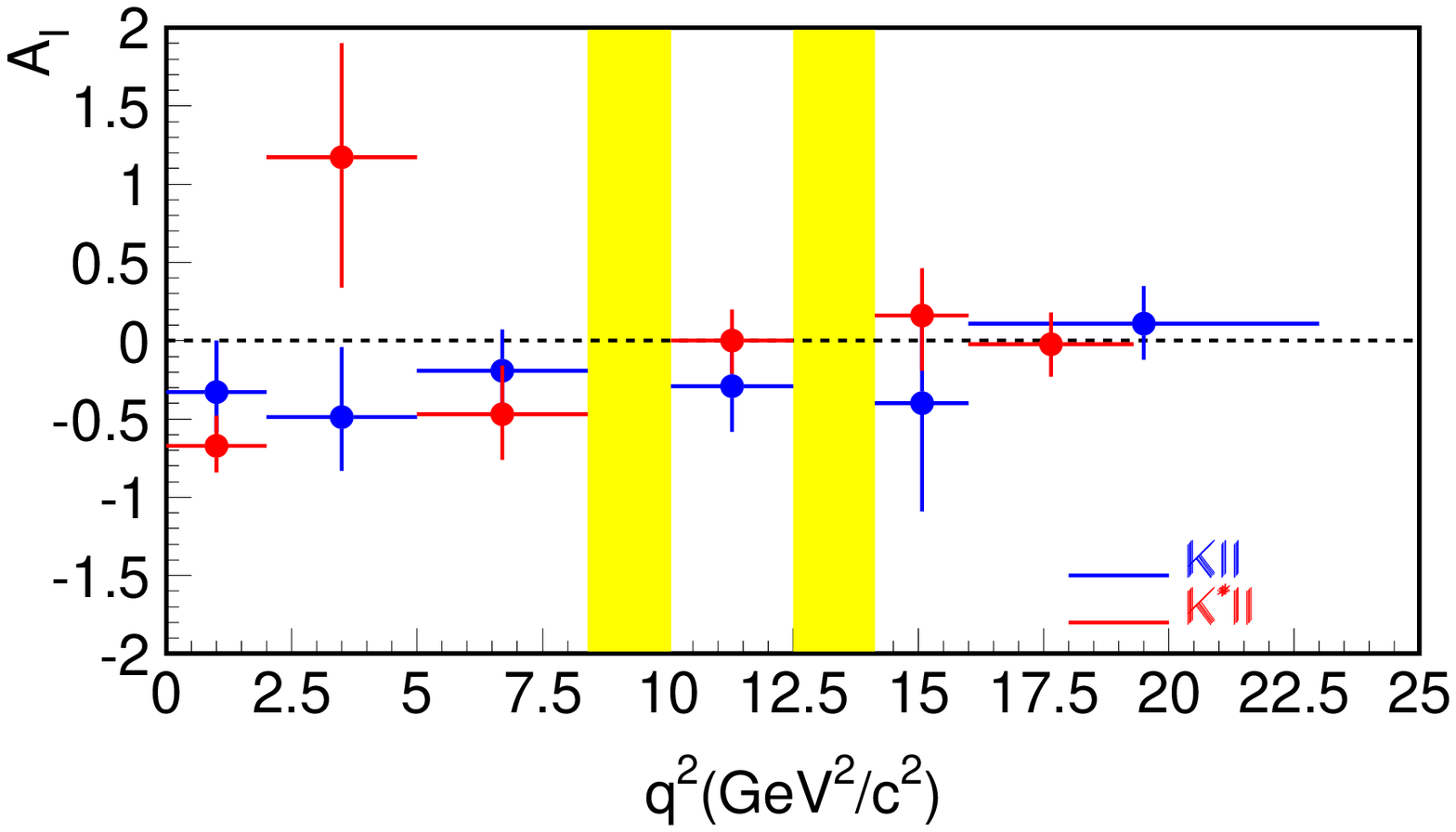}
\vskip -0.5cm
\caption{
$A_I$ as a function of $q^2$ for $K^* \ell^+ \ell^-$ (red) 
and $K \ell^+ \ell^-$ (blue) modes.}
\label{fig:ai}
\end{center}
\end{minipage}
\end{figure}

We divide $q^2$ into 6 bins and extract the signal and combinatorial background yield in each bin.
The $K^*$ longitudinal polarization fractions ($F_L$) and the forward-backward asymmetry ($A_{FB}$) are extracted from fits in the signal region 
to cos$\theta_{K^*}$ and cos$\theta_{Bl}$, respectively, where $\theta_{K^*}$ is the angle between the kaon direction and the direction opposite the $B$ meson 
in the $K^*$ rest frame, and $\theta_{Bl}$ is the angle between the $l^+(l^-)$ and the opposite of the $B(\bar{B})$ direction in the dilepton rest frame.
The differential branching fraction, $F_L$, and $A_{FB}$ as functions of $q^2$ 
for $K^{*} \ell^+ \ell^-$ and $K \ell^+ \ell^-$ modes are shown in 
Fig.~\ref{fig:dbf}, Fig.~\ref{fig:fl}, and Fig.~\ref{fig:afb}, respectively.
The differential branching fraction and $F_L$ are consistent with the Standard Model predictions.
The $A_{FB}(q^2)$ spectrum, although consistent with previous measurements~\cite{kll_exp}, 
tends to be shifted toward the positive side from the SM expectation.
A much larger data is needed for more precise measurement.
Isospin asymmetry($A_I$) is  shown in Fig.~\ref{fig:ai}. In the Standard Model,  $A_I$ is expected to be small.
Babar found a large negative asymmetry in the low $q^2$ region~\cite{bib:babar_AI}, however no significant asymmetry is found in Belle data.
%

% **************************************
\section{$b \to s \gamma$}
The decay $b \to s\gamma$ is induced through penguin diagrams.
The high energy real photon is an excellent experimental signature of the fully inclusive measurement.
\subsection{Inclusive $B \to X_s \gamma$}
\begin{figure}[h]
\vspace{-10pt}
\begin{minipage}{0.5\hsize}
\begin{center}
\includegraphics[scale=0.25]{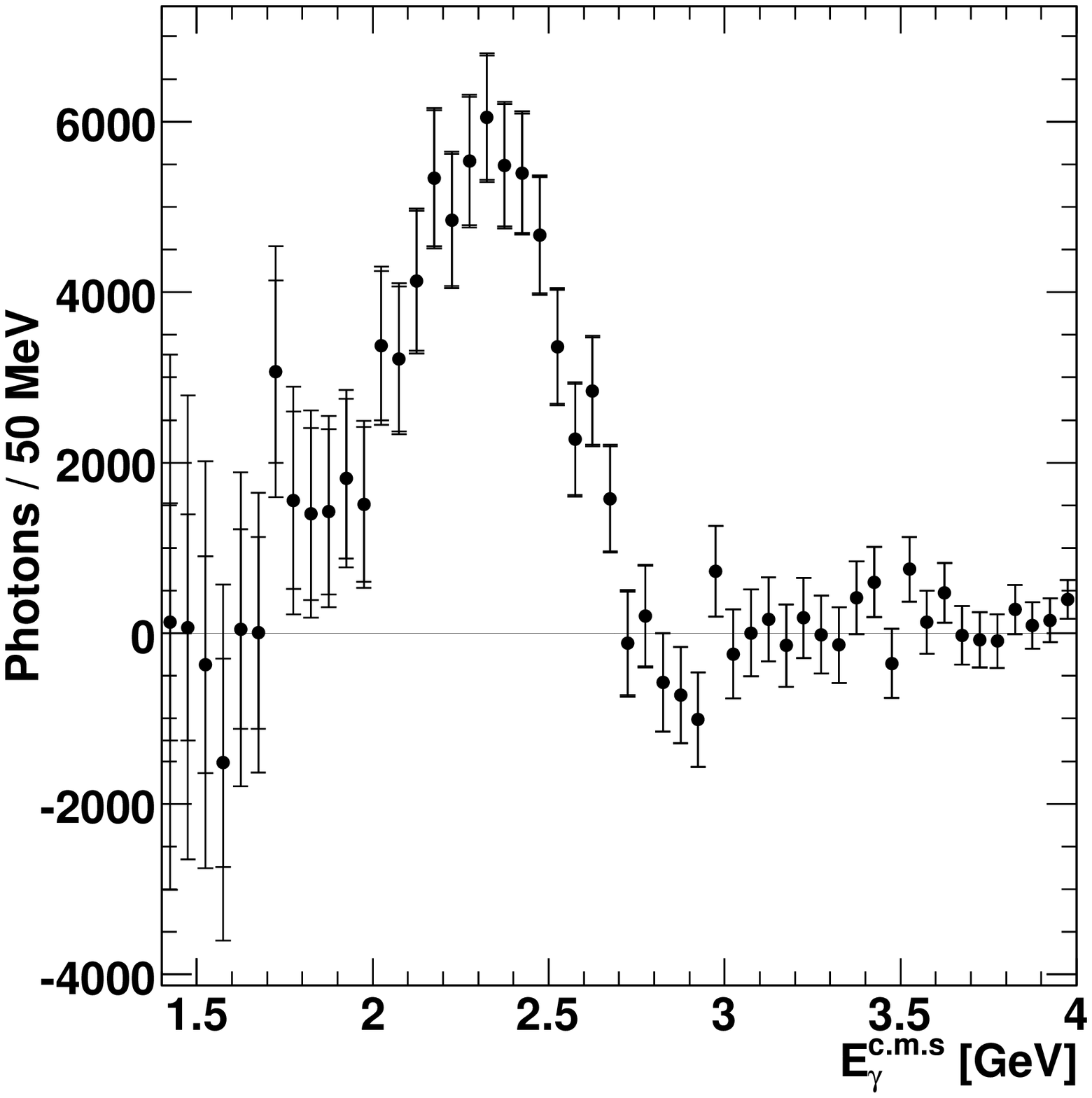} \\
  \caption{\label{fig:signal}
    The extracted photon energy spectrum of $B\to X_{s,d}\gamma$. 
    The two error bars show the statistical and total errors.
}
\end{center}
\end{minipage}
\hspace{10pt}
\begin{minipage}{0.5\hsize}
\begin{center}
\includegraphics[scale=0.35]{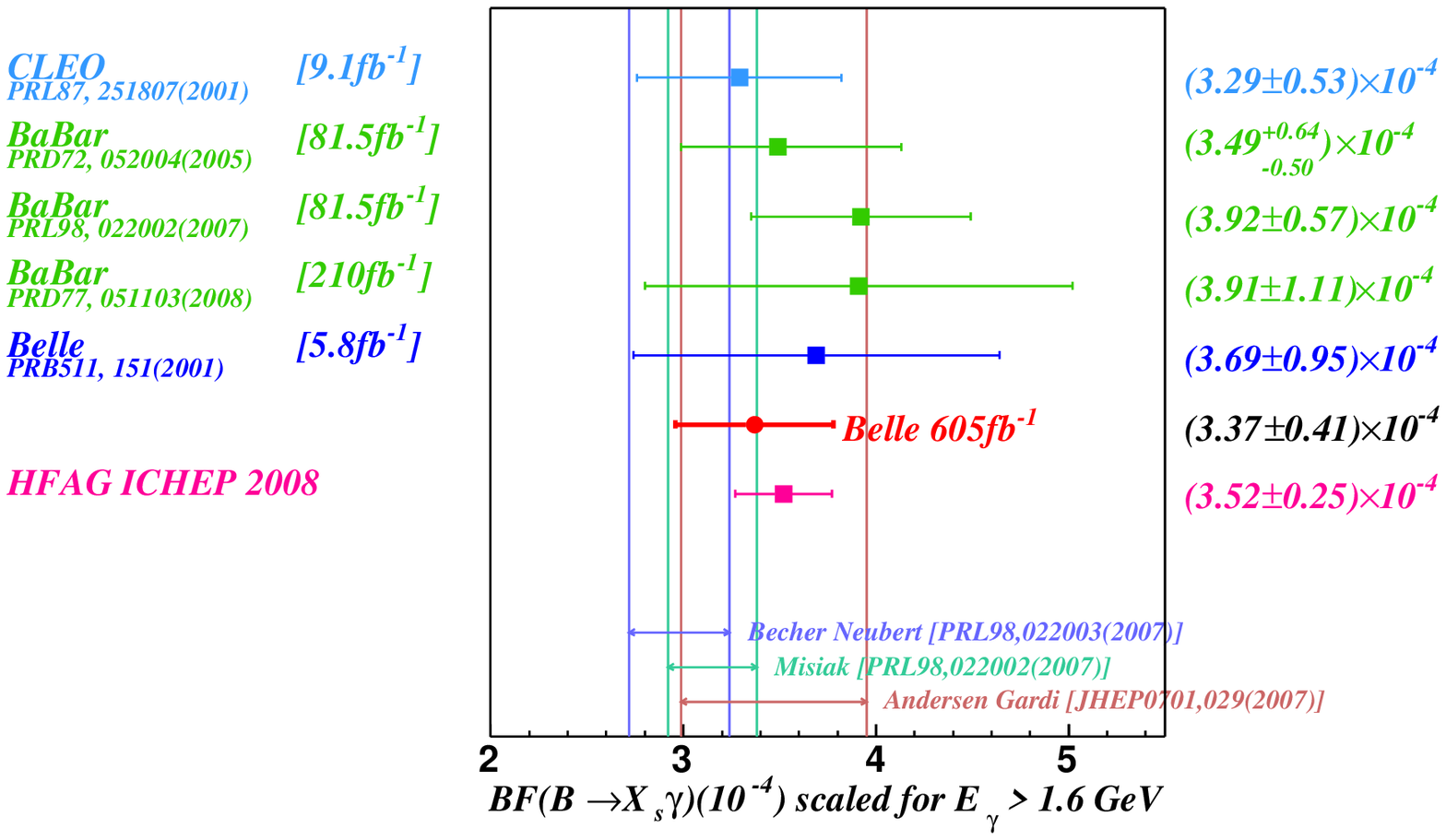} \\
  \caption{\label{fig:xsgamma_comparison}
The comparison of experimental results and theoretical predictions.
${\cal B}(B \to X_s\gamma)$ is scaled for $E^{c.m.s}_{\gamma} > 1.6 $GeV.
}
\end{center}
\end{minipage}
\end{figure}

The  ${\cal B}(B \to X_s \gamma)$ have been measured in fully inclusive method~\cite{bib:bsgamma}.
We collect  all high-energy photons, vetoing those originating from $\pi^0$ and $\eta$ decays two photons, in calorimeter.
The continuum background is suppressed using event topology information and reminder is subtracted.
We estimate the contribution from continuum event using  off-resonance data.
The events from $B$ decays are estimated using MC sample which calibrated with control data sample.
Fig.~\ref{fig:signal} show the extracted photon energy spectrum. 
We obtain ${\cal B}(B \to X_s\gamma) = (3.31 \pm 0.19 \pm 0.37 \pm 0.01) \times 10^{-4}$, 
$ \left< E_\gamma \right> = 2.281 \pm 0.032 \pm 0.053 \pm 0.002$ GeV,
$ \left <E_\gamma^2\right>-\left<E_\gamma\right>^2 = 0.0396 \pm 0.0156 \pm 0.0214 \pm 0.0012$ GeV$^2$ for $E^{c.m.s}_{\gamma} > 1.7$ GeV.
These results are the most precise measurements to date.
%
%%%%%%%%%%%%%%%%%%%%%%%%%%%%%%%%%%%%%%%%%%%%%%%%%%%%%%%%%%%%%%%%%%%%%%%%%%%
%%%%%%%%%%%%%%%%%%%%%%%%%%%%%%%%%%%%%%%%%%%%%%%%%%%%%%%%%%%%%%%%%%%%%%%%%%%%%%%%%

Fig.~\ref{fig:xsgamma_comparison} is the comparison of experimental results and theoretical predictions for the branching fraction.
The experimental results are in agreement with the theoretical predictions~\cite{bib:s_gamma_a}.
%%%%%%
\subsection{Exclusive $B \to K \eta'  \gamma$}
\begin{figure}[ht]
\centering
\includegraphics[scale=0.2]{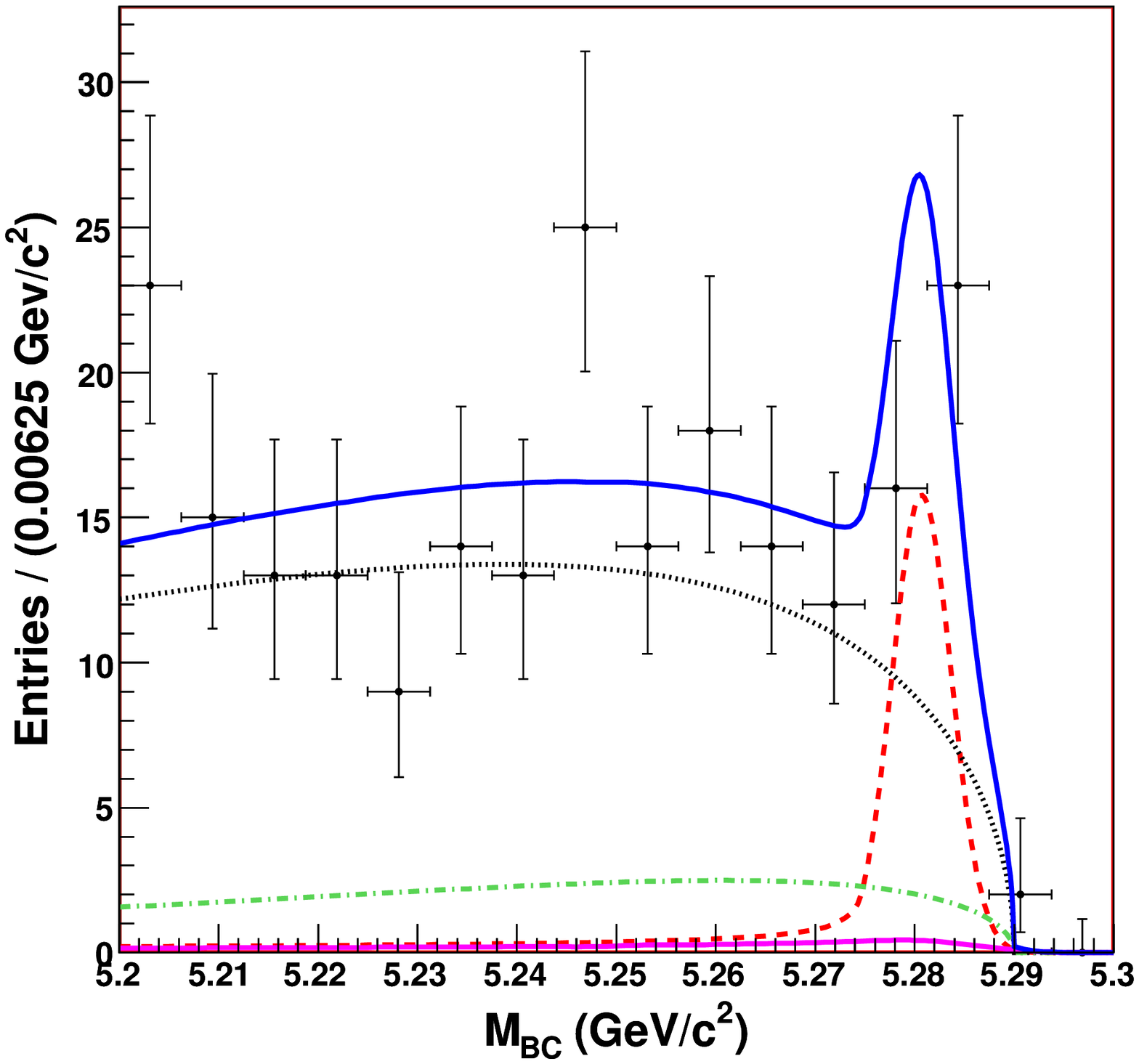}\hspace{-5pt}
\includegraphics[scale=0.2]{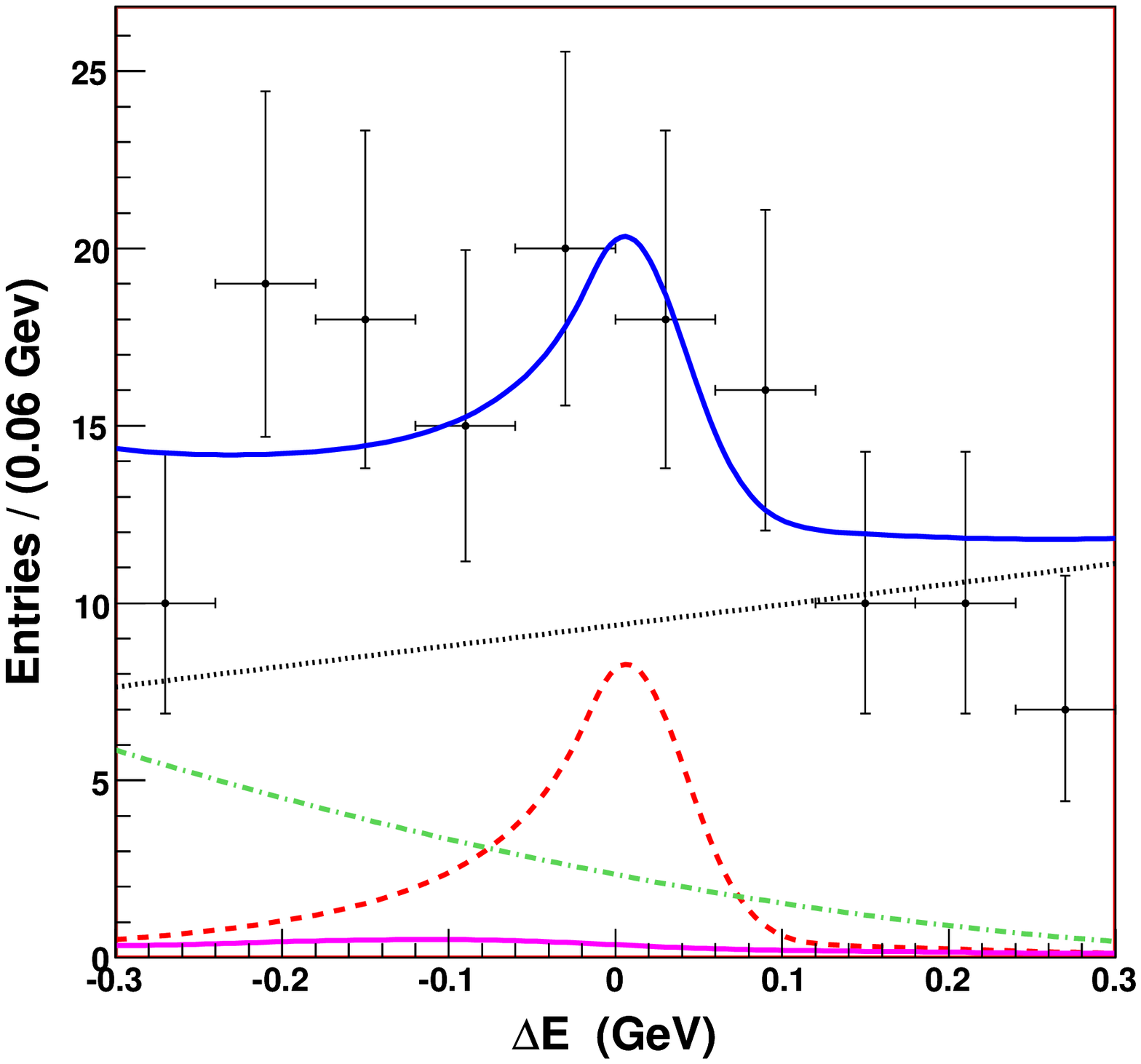}
\includegraphics[scale=0.2]{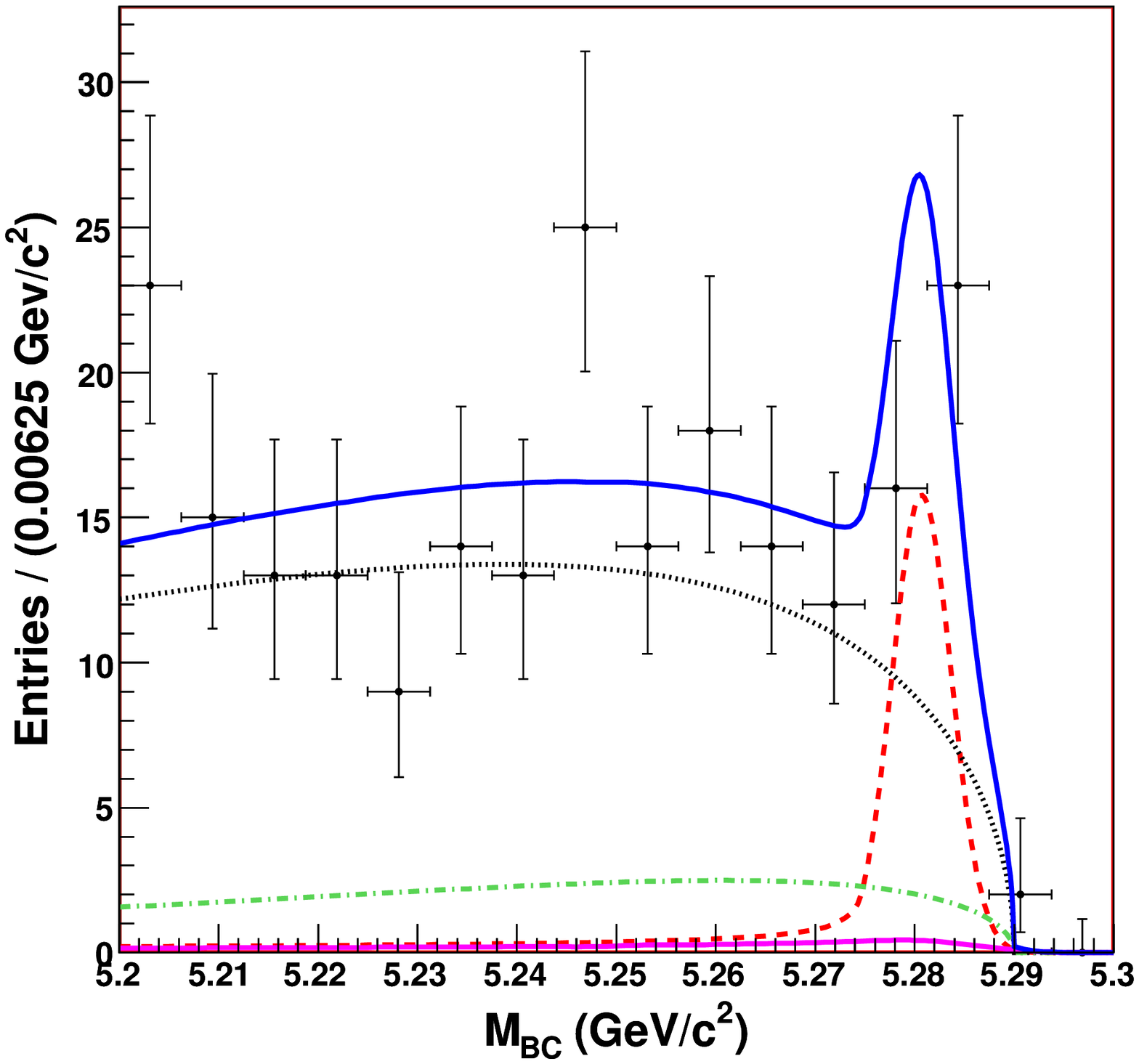}\hspace{-5pt}
\includegraphics[scale=0.2]{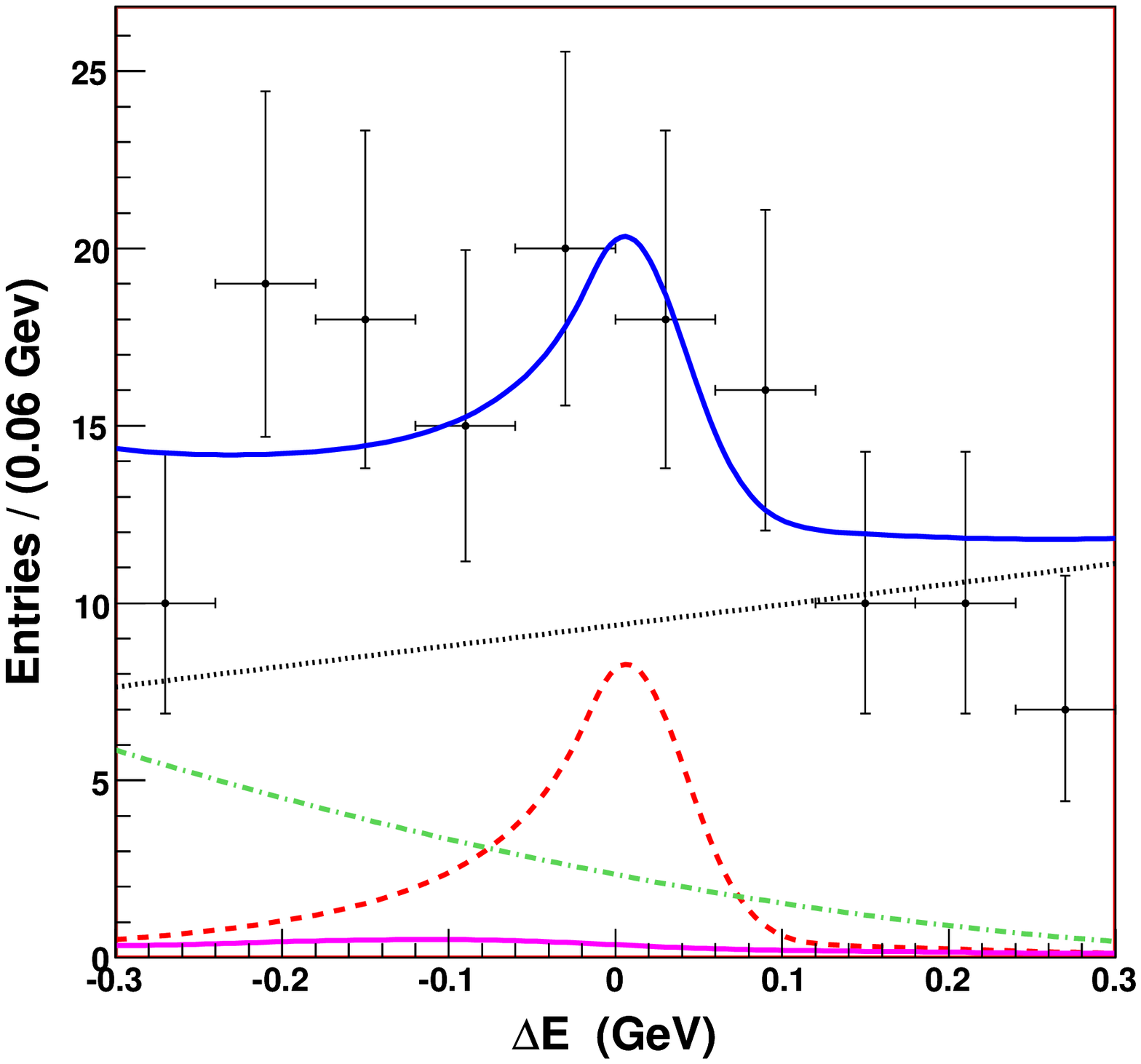}
\caption{Projections from the 2D fit to data. 
The $K\eta'\gamma$ function is shown in dashed red, 
 continuum  in dotted black, $b \to c$ in dash-dotted green, $b \to u.d.s$ in solid magenta, and th
e combined function in solid blue.  \label{fits} }
\setlength{\unitlength}{1cm}
\begin{picture}(16,0.1)
\put(0.5,5.8){{\scriptsize $B^+ \to K^+\eta'\gamma$}}
\put(4.5,5.8){{\scriptsize $B^+ \to K^+\eta'\gamma$}}
\put(8.5, 5.8){{\scriptsize $B^0 \to K^0_S \eta'\gamma$}}
\put(12.5, 5.8){{\scriptsize $B^0 \to K^0_S \eta'\gamma$}}
\end{picture}
\end{figure}

We find evidence for $B^+ \to K^+ \eta' \gamma$ decays at the $3.3\sigma$  level with 
a partial branching fraction of $(3.2^{+1.2}_{-1.1}\pm0.3)\times10^{-6}$.
  This measurement is restricted to the region of combined $K\eta'$
invariant mass less than $3.4$ GeV/$c^2$.
A 90$\%$ C.L upper limit of $6.3\times10^{-6}$ is obtained for the decay 
$B^0 \to K^0_S\eta' \gamma$ in the same $K\eta'$ invariant mass region. 
Fig.\ref{fits} shows the distributions of $M_{\rm bc}$ and $\Delta E$ with projections from 2D fit results.

\section{Summary}
We have improved measurements of differential branching fraction, isospin asymmetry, $K^*$ polarization, and forward-backward asymmetry as functions of $q^2$ for $B \to K^{(*)}ll$ decays and  branching fractions for inclusive $B\to X_s \gamma$ and the exclusive $B\to  K \eta' \gamma$ modes.
There is no evidence so far for new physics.
We need much more data sample to improve the sensitivity.  Super $B$-factory will provide one order of magnitude mode luminosity.

\end{document}